\documentclass[11pt,letterpaper]{article}
\usepackage[T1]{fontenc}
\usepackage[utf8]{inputenc}
\usepackage{lmodern}
\usepackage[margin=1in]{geometry}
\usepackage{graphicx}
\usepackage{amsmath,amssymb}
\usepackage{microtype}
\usepackage[numbers,sort&compress]{natbib}
\usepackage[font=small,labelfont=bf]{caption}
\usepackage[hidelinks]{hyperref}
\usepackage{bookmark}

\hypersetup{
    pdftitle={Hyperpolarization-enabled cooperative molecular magnetic amplification},
    pdfauthor={Jingyan Xu, Raphael Kircher, Min Jiang, Xinhua Peng, Danila A. Barskiy, Dmitry Budker}
}
\begin{document}
\begin{center}
{\fontsize{18}{22}\selectfont\bfseries Hyperpolarization-enabled cooperative molecular magnetic amplification\par}
\vspace{0.7em}
{\normalsize Jingyan Xu\textsuperscript{1,2,3},
Raphael Kircher\textsuperscript{1,2,3},
Min Jiang\textsuperscript{6,7,8},
Xinhua Peng\textsuperscript{6,7,8},\\[0.2em]
Danila A. Barskiy\textsuperscript{1,2,3,5,*},
Dmitry Budker\textsuperscript{1,2,3,4}\par}
\vspace{0.5em}
{\fontsize{8.5}{10.5}\selectfont
\textsuperscript{1}Helmholtz Institute Mainz, Mainz, 55128, Germany.\par
\textsuperscript{2}GSI Helmholtzzentrum f\"ur Schwerionenforschung, Darmstadt, 64291, Germany.\par
\textsuperscript{3}Institute of Physics, Johannes Gutenberg-Universit\"at, Mainz, 55099, Germany.\par
\textsuperscript{4}Department of Physics, University of California, Berkeley, CA, 94720, USA.\par
\textsuperscript{5}Frost Institute for Chemistry and Molecular Science, Department of Chemistry, Department of Physics,
University of Miami, Coral Gables, FL, 33146, USA.\par
\textsuperscript{6}Laboratory of Spin Magnetic Resonance, School of Physical Sciences, Anhui Province Key Laboratory of Scientific Instrument Development and Application,
University of Science and Technology of China, Hefei, 230026, China.\par
\textsuperscript{7}Hefei National Laboratory, University of Science and Technology of China, Hefei, 230088, China.\par
\textsuperscript{8}Hefei National Research Center for Physical Sciences at the Microscale,
University of Science and Technology of China, Hefei, 230026, China.\par
\vspace{0.4em}
\textsuperscript{*}Corresponding author: \href{mailto:barskiy@miami.edu}{\nolinkurl{barskiy@miami.edu}}\par
Other author contacts: \nolinkurl{jixu@uni-mainz.de}; \nolinkurl{rkircher@uni-mainz.de};
\nolinkurl{dxjm@ustc.edu.cn}; \nolinkurl{xhpeng@ustc.edu.cn}; \nolinkurl{dbukder@uni-mainz.de}.\par
}
\end{center}
\begin{abstract}
Detecting weak magnetic fields at hertz frequencies is central to biomagnetism, geophysical sensing, and searches for new physics, yet it remains challenging because of technical systematics and environmental noise. Here, we realize a continuously operating molecular magnetic amplifier. Signal amplification by reversible exchange (SABRE) parahydrogen-based hyperpolarization maintains non-equilibrium spin order at zero field, turning scalar-coupled nuclear spins in liquid-state molecules into an active, frequency-selective gain medium. Measurements of its complex magnetic susceptibility ($\chi$) reveal an intrinsic response of $|\chi|\approx10$. By feeding the collectively generated magnetic field back to the spins below the threshold for self-sustained $J$-oscillation, we extend the effective spin coherence and cooperatively enhance the susceptibility up to $|\chi|\approx80$. Near threshold, this produces approximately fourfold amplification of a weak (20\,pT) magnetic field at 3.37\,Hz at the sensor location. These results establish hyperpolarized molecules as a tunable active magnetic medium for narrowband quantum sensing and precision measurement.
\end{abstract}
{\small\noindent\textbf{Keywords:} hyperpolarization, magnetic amplification, nuclear magnetic resonance, active feedback, quantum sensing.\par}

\section{Introduction}
Nuclear spins combine long coherence times with chemically specific
interactions and are ubiquitous in matter, making them indispensable
to chemical analysis, medical imaging, and materials characterization \cite{ernst1990principles, budinger1984nuclear, levitt2008spin}.
These same properties also make nuclear spins attractive quantum sensors
for precision magnetometry and searches for weak spin-dependent
interactions. Performance of nuclear spins as sensing systems, however, is
fundamentally constrained by their small gyromagnetic ratios and the
low spin polarization available at thermal equilibrium \cite{ardenkjaer2015facing, barskiy2025zero}. Hyperpolarization offers a route to overcoming this limitation by preparing molecular spin ensembles far from thermal equilibrium and, thus, increasing their field response by several orders of magnitude \cite{ardenkjaer2015facing}. In this regime,
nuclear spins can serve not only as passive spectroscopic probes but also
as an active medium able to coherently amplify weak magnetic signals \cite{jiang2021search,jiang2024observation,zhou2026hyperpolarized,jiang2022floquet,jiang2021floquetmaser,su2022review}.

\par Zero-field, scalar-coupled molecular spin systems are particularly attractive for low-frequency magnetic amplification. In the absence of Zeeman splitting,
their transition frequencies are determined primarily by intramolecular $J$-couplings and naturally occur in the hertz range \cite{barskiy2025zero}. Weak magnetic signals in this frequency range arise in biomagnetic measurements, including magnetoencephalography \cite{baillet2017magnetoencephalography,rhodes2023measurement}, magnetocardiography \cite{cohen1970magnetocardiograms,yang2021new}, and in geophysical phenomena
such as Pc1 geomagnetic pulsations \cite{jacobs1964classification} and Schumann resonances \cite{SR_science}. Oscillatory magnetic and pseudomagnetic fields in the hertz regime are also relevant to searches for ultralight dark matter \cite{jiang2021search}. Detecting such signals is challenging because low-frequency measurements are often limited by technical drift and environmental magnetic noise. A continuously operating, phase-preserving amplifier could therefore enhance a target signal before
readout. Molecular spin ensembles provide intrinsic frequency selectivity for this purpose because their resonance frequencies and linewidths are governed
by microscopic spin interactions and can, in principle, be tuned through molecular structure, isotopic composition, choice of solvent, and applied bias fields.

\par Signal amplification by reversible exchange (SABRE) with parahydrogen provides an efficient means of hyperpolarizing $J$-coupled molecules \textit{in situ} in the liquid phase at zero magnetic field
\cite{adams2009reversible,theis2012zero}. SABRE can therefore supply the nonequilibrium spin order required to turn low-frequency molecular $J$-transitions into an active magnetic gain medium. Moreover, continuous
parahydrogen exchange can replenish this spin order during measurement, offering a route from transient amplification to continuously pumped
operation.

\par Previous work demonstrated magnetic amplification using hyperpolarized molecular nuclear spins,  where SABRE-hyperpolarized molecular spin ensembles generated a measurable magnetic response to weak oscillating fields \cite{zhou2026hyperpolarized}. That experiment established the basic principle that the hyperpolarized spin ensemble can generate a dipolar field in response to an applied magnetic field, thereby enabling amplification of weak ($<$1\,nT) magnetic signals \cite{barskiy2025zero}. However, the amplification was realized in a transient mode and lasted only over the coherence lifetime of the initially hyperpolarized spins. Moreover, an anomalous phase dependence of the magnetic amplification was observed and it was attributed to interference between the applied and molecule-generated fields \cite{zhou2026hyperpolarized}. The response was not formulated in terms of a calibrated complex magnetic susceptibility, leaving the intrinsic molecular response and its relation to the geometry-dependent amplification measured at the sensor unresolved.  In addition, the reported magnetic amplification remained relatively modest, with a maximum gain at the position of the sensor of approximately 13\% \cite{zhou2026hyperpolarized}.

\par In this letter, we address these problems by realizing a continuous magnetically active molecular gain medium based on SABRE-hyperpolarized scalar-coupled spins. Continuous parahydrogen bubbling maintains a non-equilibrium population imbalance across the molecular $J$-coupled transitions, turning the liquid sample into an active magnetic gain medium. When a weak input oscillating field is applied near a molecular $J$-transition, the driven ensemble is excited and generates a macroscopic oscillating magnetization. The nearby magnetometer therefore detects the output magnetic field consisting of the applied input field, $B_a(t)$, and the sample-generated dipolar field, $B_s(t)$ (Fig.\,\ref{fig:schematics}a).

\par A central feature of this amplifier is that its gain is governed by the complex magnetic susceptibility of the hyperpolarized molecular medium, which determines both the amplitude and phase of the induced magnetization in response to the applied input field.
By directly measuring the complex susceptibility, we quantitatively separate the intrinsic molecular response from the geometry- and phase-dependent amplification measured at the sensor and describe the observed dispersive amplification in terms of interference between the applied field and the phase-shifted molecular field. More generally, the measured magnetic susceptibility is connected to the molecular parameters including spin density, polarization, coherence lifetime, and zero-field $J$-transition structure. As a result, in addition to quantifying the magnetic amplification of the gain medium, continuous measurement of the sample’s magnetic susceptibility provides an alternative approach to spectroscopy with direct access to rich molecular information. In contrast to conventional free-decay zero-field NMR, where the acquisition window is limited by short-lived spin coherences, continuous susceptibility measurements enable sustained observation of molecular spectral features.

\par To overcome the limited magnetic amplification of the molecular medium, we introduce external active feedback to realize the so-called cooperative magnetic amplification, following related feedback-enabled cooperative spin amplification in noble-gas ensembles \cite{xu2024cooperative,jiang2025interacting}. Building on the external programmable feedback loop previously developed for zero-field quantum $J$-oscillators \cite{xu2026quantum}, the collective signal generated by the hyperpolarized spin ensemble is detected with an optically pumped magnetometer (OPM), isolated from the applied field through gradiometric measurement \cite{blanchard2020zero}, and then fed back with a controlled external gain $g$ and feedback delay $t_d$, as illustrated in Fig.\,\ref{fig:schematics}a. This feedback is also applied continuously during the measurement sequence shown in Fig.\,\ref{fig:schematics}b, coupling the molecular nuclear spins through their collectively generated field. In the presence of active feedback, the output magnetic field is reconstructed by subtracting the applied feedback field from the raw OPM signal using the known feedback-control voltage and the calibrated coil response. When operated with feedback gain $g$ below the threshold for spontaneous $J$-oscillations \cite{xu2026quantum}, this cooperative interaction extends the effective coherence lifetime and enhances the magnetic response of the hyperpolarized spin amplifier. Operating near the threshold, we demonstrate approximately fourfold magnetic amplification near the $2J$-resonance, substantially exceeding the amplification achieved without active feedback.
The same amplification scheme can be applied to the $J$-transition by matching the feedback delay value to excite the $J$-resonance.

\section{Continuous gain from hyperpolarized \texorpdfstring{$J$}{J}-coupled spins}
\begin{figure}[tbp]
    \centering
\includegraphics[width=\linewidth]{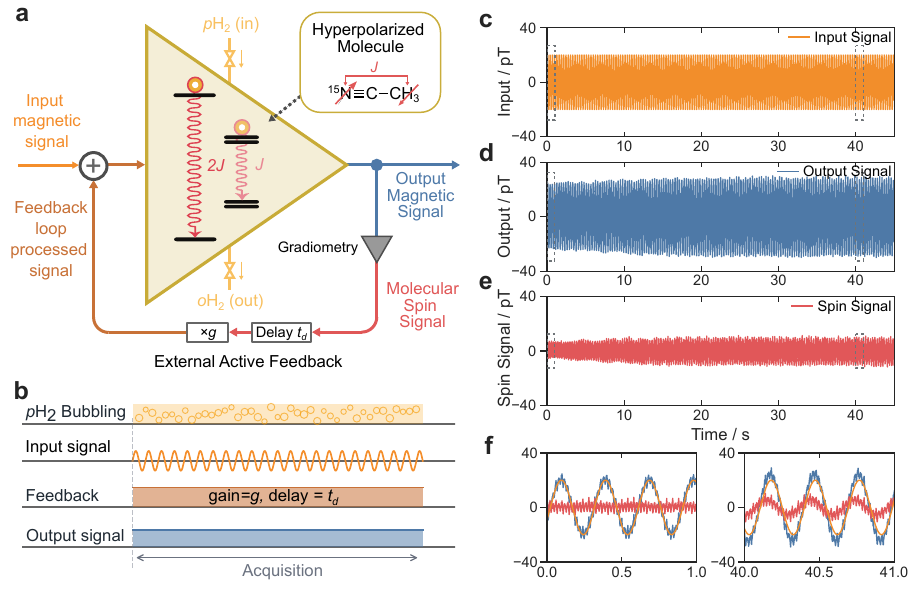}
    \caption{(a) Schematic of the molecular magnetic amplifier. SABRE-hyperpolarized [$^{15}$N]-acetonitrile at zero magnetic field serves as an active gain medium, amplifying weak magnetic inputs through population-inverted $J$-coupled spin transitions. To realize cooperative molecular amplification, the molecular spin signal is extracted through gradiometric detection and fed back to the sample via an external feedback loop, which amplifies and delays the molecular spin signal in real time. This active feedback scheme effectively extends the spin coherence time, thereby enhancing the magnetic response and amplification gain of the molecular medium. (b) Experimental sequence used to characterize the performance of the molecular amplifier. $p$H$_2$ bubbling, application of a test input magnetic signal, external feedback, and signal acquisition are performed continuously.
    (c) Input magnetic signal (orange), reconstructed from the applied voltage and normalized to magnetic-field units ($f=3.393$\,Hz).
    (d) Output magnetic signal measured by the optically pumped magnetometer (OPM) in the absence of active feedback ($g=0$) (blue).
    (e) Molecular-spin-generated dipolar field at the OPM position (red), obtained by subtracting the input signal in (c) from the measured signal in (d).
    (f) Zoomed-in views of traces in (c)--(e) at the beginning of the sequence (0--1\,s) and at steady state (40--41\,s). High frequency signal is associated with power-line 50\,Hz.
}
    \label{fig:schematics}
\end{figure}

We experimentally realize a magnetically active molecular gain medium using a liquid sample consisting of 5\% isotopically labeled [$^{15}$N]-acetonitrile mixed with 95\% [$^{14}$N]-acetonitrile-$d_3$. The present SABRE chemistry requires a neat acetonitrile environment; $[^{14}\mathrm{N}]$-acetonitrile-$d_3$ preserves this chemical environment while reducing the proton density and, consequently, intermolecular dipole--dipole relaxation of $[^{15}\mathrm{N}]$-acetonitrile. In addition, it introduces no overlapping spectral features in the frequency range of interest. The sample is contained in an NMR tube with a sphere at the end (inner diameter $\sim$14.5 mm), as illustrated in Fig.\,\ref{fig:schematics}a \cite{xu2025situ}. In $^{15}\mathrm{N}$-acetonitrile at zero magnetic field, the relevant
transitions occur near $1.69\,\mathrm{Hz}$ ($J$) and
$3.37\,\mathrm{Hz}$ ($2J$), with the latter providing the operating frequency for the magnetic amplification demonstrated in this work.

\par Bubbling of parahydrogen (\textit{p}H$_2$) through the solution containing the isotopic mixture and an activated iridium-based SABRE catalyst enables the continuous polarization transfer from \textit{p}H$_2$ to [$^{15}$N]-acetonitrile molecules. At zero magnetic field, this process generates a population imbalance among the molecular scalar-coupling ($J$) transitions, as shown in Fig.\,\ref{fig:schematics}a \cite{theis2012zero,xu2025situ}. When a weak oscillating magnetic field is applied near resonance with a $J$ or $2J$ transition, it drives the SABRE-generated population imbalance into coherences between the corresponding $J$-transitions. These coherences correspond to a time-dependent macroscopic magnetization of the molecular ensemble, which in turn generates an oscillating dipolar magnetic field detectable by the nearby OPMs. Consequently, the hyperpolarized ensemble behaves as an active magnetic gain medium for the magnetic fields near-resonant with $J$-transitions.

\par The operation of the molecular amplifier is characterized using the continuous measurement protocol illustrated in Fig.\,\ref{fig:schematics}b. Throughout the experiment, parahydrogen bubbling, application of a weak test magnetic field through the magnetic shield’s shimming coil, signal acquisition, and, when enabled, electronic feedback are performed continuously. We first consider the response of the active medium in the absence of feedback ($g=0$). A representative measurement is shown in Fig.\,\ref{fig:schematics}c–f. Figure\,\ref{fig:schematics}c shows the applied input magnetic field, reconstructed from the applied voltage using the calibrated coil response, while Fig.\,\ref{fig:schematics}d shows the total magnetic field measured with the OPM. By subtracting the known input field from the measured signal, we isolate the magnetic field generated by the hyperpolarized molecular ensemble, shown in Fig.\,\ref{fig:schematics}e. This molecular-spin-generated dipolar field arises from the driven transition coherences described above and represents the intrinsic response of the active medium to the external field. The temporal evolution of the input field, measured field, and molecular response, both at the onset of excitation and after the system reaches steady state, is shown in the zoomed-in views of Fig.\,\ref{fig:schematics}f.

\section{Complex susceptibility of the gain medium}
To characterize the intrinsic response of the hyperpolarized molecular gain medium, we apply a weak oscillating magnetic field with amplitude 200 pT and sweep its frequency across the relevant $J$-transitions. The measured response is quantified through the complex magnetic susceptibility, $\chi(\omega)$, which relates the induced molecular magnetization, $M(t)$, to the applied input magnetic field through
\begin{equation}\label{eq:chi}
    \mu_0 M(t)=\mathrm{Re}[\chi(w) B_a e^{iwt}],
\end{equation}
where $B_a$ is the amplitude of the input magnetic field and $\mu_0$ is the vacuum permeability; note that the complex susceptibility $\chi(\omega)$ is dimensionless. Thus, $|\chi(\omega)|$ quantifies the molecule-generated field within the sample relative to the
applied field, independently of the detection geometry. The OPM does not measure
this field directly. At the sensor position, the molecular field is reduced by
the geometric coupling factor $\kappa=2r^3/(3d^3)$, where $r$ is the radius of the spherical sample and $d$ is the center-to-center distance between the sample and the OPM cell. In our setup, $r=6.25$\,mm and $d=15.45$\,mm, giving $\kappa=0.044$. Experimentally, the complex susceptibility is reconstructed from the amplitude
and phase of the molecular-spin-generated field measured by the OPM relative
to the applied excitation. The extracted real and imaginary parts of the susceptibility are shown in Fig.\,2a-b. Two resonances are observed near the $J$- and $2J$-transition frequencies, corresponding to the $K=1/2$ and $K=3/2$ manifolds (i.e., two possible realizations of the total hydrogen spin) of [$^{15}$N]-acetonitrile, respectively.

\par In the weak driving regime (i.e., $ | B_a (\gamma_{^1\rm H}-\gamma_{^{15}\rm N}) | \ll  \Gamma_{K}$), where $\chi(w)$ is independent of the input field amplitude,
the susceptibility is described by a two-pole response function:
\begin{equation}
\chi(\omega)=\frac{1}{\kappa}\left[
\frac{g_{1/2}\Gamma_{1/2}}{(2\pi J-\omega)+i\Gamma_{1/2}} + \frac{g_{3/2}\Gamma_{3/2}}{(4\pi J-\omega)+i\Gamma_{3/2}} \right]\, ,
\end{equation}
where $g_K$ and $\Gamma_K$ denote the internal gain factor and relaxation rate of the corresponding spin manifold, respectively. The internal gain factors $g_{K}$
(specifically, $g_{1/2}= 0.5 \kappa
\mu_0 \hbar [C]N_A
(\gamma_{\rm ^1 H}-\gamma_{\rm ^{15} N})^2  {\Gamma_{1/2}^{-1}} P_{\rm ss} $ and $g_{3/2}= 1.25 \kappa
\mu_0 \hbar [C]N_A
(\gamma_{\rm ^1 H}-\gamma_{\rm ^{15} N})^2  {\Gamma_{3/2}^{-1}} P_{\rm ss}$) are defined as the ratio between the sample-generated magnetic field measured by the OPM and the applied field at the corresponding $J$- and $2J$-resonances. Here, $K=1/2$ and $3/2$ label the manifolds associated with the $J$- and $2J$-resonances, respectively; $[C]$ is the molecular concentration; $N_A$ is Avogadro's constant; $P_{\mathrm{ss}}$ is the steady-state scalar spin order; $\Gamma_K$ is the relaxation rate of the corresponding manifold; $\gamma_{{}^{1}\mathrm{H}}$ and
$\gamma_{{}^{15}\mathrm{N}}$ are the gyromagnetic ratios of
$^{1}\mathrm{H}$ and $^{15}\mathrm{N}$; and $\mu_0$ and $\hbar$ are the vacuum permeability and the reduced Planck's constant, respectively. Solid lines in Fig.\,2a-b show simultaneous fits to the real and imaginary components of the measured susceptibility, yielding excellent agreement with the data. The dispersive lineshapes in $\mathrm{Re}[\chi]$ and absorptive peaks in $\mathrm{Im}[\chi]$ are characteristic of a resonantly driven gain medium. The fact that the imaginary component of susceptibility is a positive number at resonance indicates that the internal gain as well as steady-state scalar spin order ($P_{\rm ss}$) are both negative.

\par To probe the limits of the linear-response regime, we measure the molecular-spin-generated field as a function of input magnetic-field amplitude at the $2J$-resonance. As shown in Fig.\,2c, the response initially increases linearly with the drive amplitude, as expected from the field-independent susceptibility in the weak-driving regime. Experimentally, the response remains linear for $B_a<250\,$pT, corresponding to  $ | B_a (\gamma_{^1\rm H}-\gamma_{^{15}\rm N}) | < \Gamma_{3/2} / 10 $ near resonance. At stronger drive, however, the response becomes nonlinear, reaches a maximum, and eventually decreases. This behavior arises from saturation of the driven transition: as the driving strength becomes comparable to the relaxation rate, the SABRE-maintained population imbalance is progressively depleted and less spin order remains available for conversion into coherent magnetization. The solid curve is calculated using the nonlinear extension of the susceptibility model, with the parameters $g_{3/2}$ and $\Gamma_{3/2}$ fixed to the values extracted from the susceptibility measurements in Fig.\,2a-b.

\par
\begin{figure}[tbp]
    \centering
\includegraphics[width=0.68\linewidth]{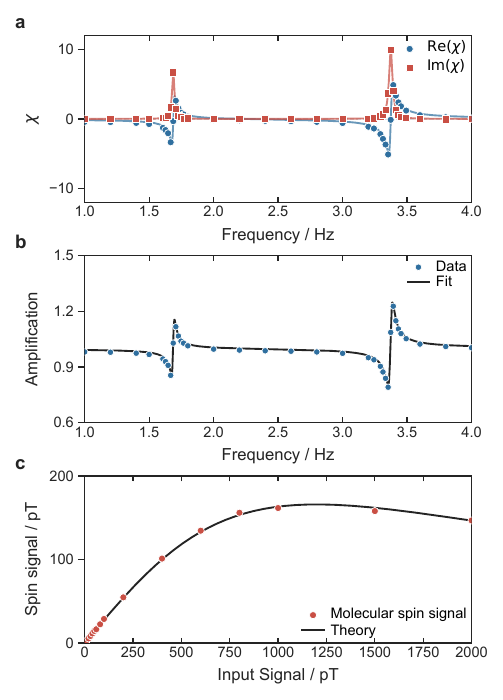}
    \caption{
    Frequency response of the magnetically active molecular gain medium in the absence of external feedback ($g=0$). (a) Real (blue circles) and imaginary (red squares) components of the magnetic susceptibility $\chi$ as a function of applied-field frequency, measured using a 200\,pT oscillating magnetic field. The complex susceptibility is extracted from the amplitude and phase of the molecular-spin-generated field measured by the OPM relative to the applied excitation. (b) Corresponding frequency-dependent amplification.  Markers denote experimental data; solid curves show theoretical predictions using parameters calibrated from the susceptibility fit in (a). (c) Amplitude of the molecular-spin-generated dipolar field at the OPM position as a function of applied-field amplitude, measured at a fixed frequency of 3.372\,Hz near the 2$J$-reasonance. The response is linear at weak drive but becomes sublinear and eventually decreases for input fields $\sim$1\,nT and above because strong driving saturates the transition and reduces the steady-state population imbalance available for conversion into spin coherence. The solid curve shows the nonlinear steady-state theory, using parameters determined from the susceptibility measurement in (a).}
    \label{fig:fig2}
\end{figure}

 The experimentally measured susceptibility determines the magnetic amplification produced by the molecular ensemble. Specifically, the amplification factor is related to the susceptibility through
\begin{equation}\label{eq:Ag0}
A(\omega)=\left|1+\kappa\chi(\omega)\right|\,.
\end{equation}
This distinction between intrinsic susceptibility and measured amplification is important. In the absence of feedback, at resonance, the susceptibility reaches
$|\chi(\omega)|\simeq10$ (Fig.\,\ref{fig:fig2}a), indicating that the characteristic molecule-generated field within the sample is approximately one order of magnitude larger than the applied field. In the present geometry, however, $\kappa=0.044$, so only a maximum molecular-field contribution of approximately
$\kappa|\chi|\simeq0.44$ reaches the OPM. The field detected by the OPM is the sum of this molecule-generated field and the applied field;
therefore, the measured amplification depends on both the magnitude and phase of the complex susceptibility.

\par The experimentally measured amplification spectrum, obtained as the ratio between the output and input magnetic-field amplitudes, is shown in Fig.\,\ref{fig:fig2}c. The solid curves are calculated from Eq.(\ref{eq:Ag0}) using the susceptibility extracted from Fig.\,2a-b, and are in excellent agreement with the measured amplification. For $|\kappa\chi|\ll1$, the amplification reduces to $A(\omega)\simeq1+\kappa\mathrm{Re}[\chi(\omega)]$, giving rise to the observed dispersive frequency dependence. At exact resonance,
$\mathrm{Re}[\chi]=0$ and the molecular response is in quadrature with the applied field, so the amplification remains close to unity. Away from resonance, the in-phase component of the molecular response produces constructive or destructive interference with the applied field. Consequently, the passive amplification reaches approximately 1.4 at finite detuning, despite the substantially larger intrinsic field generated within the sample.

\section{Feedback extends the effective spin coherence lifetime}
The susceptibility of the gain media and, therefore, the magnetic amplification depend on both the hyperpolarization level and the coherence lifetime of the molecular ensemble. Whereas the spin polarization is set by the SABRE pumping process, the coherence lifetime can be extended through active feedback.

\par To quantify this effect, we measure the effective relaxation rate under active feedback using the protocol shown in Fig.\,3a. A weak oscillating magnetic field (200\,pT at 3.372\,Hz) is first applied to selectively excite the $2J$ transition with a small nutation angle. The excitation field is then switched off and the active feedback is turned on with a set feedback delay time of $t_d=222\,\mathrm{ms}$. The subsequent free decay of the molecular-spin-generated magnetic field is recorded by the OPM, with the applied feedback field actively subtracted from the measured signal. Representative decay traces for several feedback gains are shown in Fig.\,3b. As the magnitude of the negative feedback gain is increased, the decay becomes slower, indicating a reduction of the effective relaxation rate.

\par
\begin{figure}[tbp]
    \centering
\includegraphics[width=0.68\linewidth]{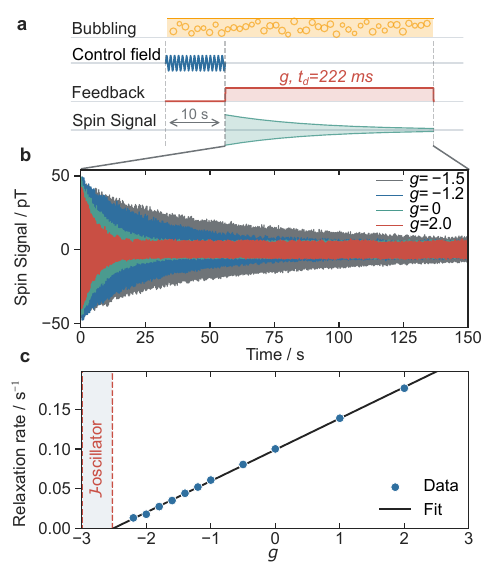}
    \caption{(a) Experimental sequence used to characterize the effective relaxation rate of the molecular spins under active feedback. A weak oscillating magnetic field (200\,pT at 3.372\,Hz) is applied to selectively excite the $2J$ transition in the small-nutation-angle regime, followed by signal acquisition under the active feedback scheme. (b) Time-domain traces of the molecular-spin-generated dipolar field recorded by the OPM sensor for feedback gains $g=0$, $-1.2$, and $-1.5$. The feedback delay is fixed at $\tau=222$\,ms, corresponding to a $3\pi/2$ phase delay for the $2J$ transition. (c) Effective relaxation rates extracted from the data in (b) using decaying-sinusoid fits. Solid lines show linear fits. When the feedback gain exceeds the threshold value, the system runs self-sustained spontaneous $J$-oscillations.}
    \label{fig:fig3}
\end{figure}

 The effective relaxation rates extracted from decaying-sinusoid fits are plotted in Fig.\,3c. For the experimentally chosen delay $t_d=222\,\mathrm{ms}$, corresponding to a phase delay of approximately $3\pi/2$ at the $2J$ resonance, the feedback primarily modifies the effective relaxation rate while leaving the resonance frequency nearly unchanged. The effective relaxation rate $\Gamma_{\rm eff}$ is expected to decrease linearly with feedback gain and approach zero at the threshold for the $J$-oscillation to happen,
\begin{equation}\label{eq:g_eff}
    \Gamma_{\rm eff} =  \Gamma_{3/2}(1 + g_{3/2}\,g) \, ,
\end{equation}
where $g_{3/2}$ is the internal gain defined above and $g$ is the
externally applied feedback gain. The measured relaxation rates are
well described by this linear dependence [Fig.\,\ref{fig:fig3}(c)].
At the threshold $g=-1/g_{3/2}$, the effective relaxation rate
vanishes. Beyond this threshold, the feedback overcomes the intrinsic
relaxation, giving rise to self-sustained $J$-oscillations without an
initial excitation \cite{xu2026quantum}.

\section{Cooperative magnetic amplification}
The increase of the molecular ensemble's effective coherence is expected to directly enhance its response to the input magnetic field and, consequently, the magnetic amplification. To probe this effect, we apply a weak oscillating magnetic field with an amplitude of 20\,pT, ensuring operation within the linear-response regime, defined empirically by the criterion $| B_a(\gamma_{^{1}\rm H}-\gamma_{^{15}\rm N}) |<0.1 \, \Gamma_{\rm eff}$ based on the saturation measurements shown in Fig.\,2c. In addition, we maintain the feedback $t_d=222$\,ms, making sure an approximately $ 3 \pi / 2$ phase delay for the $2J$ transition.

\par As an example, Figure\,4a shows the reconstructed output magnetic field for an input field of 20\,pT applied at $f=3.374$\,Hz and a feedback gain of $g=-2.2$, obtained after subtracting the applied feedback field from the OPM signal. At the beginning of the sequence, the measured output is dominated by the applied input field. Over time, the molecule-generated magnetic field builds up and becomes the dominant contribution to the output signal, leading to a strongly amplified steady-state response.

\par To quantify the feedback-induced modification of the molecular response, Figs.\,4c and 4d show the corresponding real and imaginary components
of the magnetic susceptibility for several feedback gains. Increasing the magnitude of the feedback gain reduces the effective relaxation rate and consequently enhances the magnetic susceptibility near resonance. The
dispersive $\mathrm{Re}[\chi]$ and absorptive $\mathrm{Im}[\chi]$ lineshapes are preserved, while their linewidths progressively narrow and their peak amplitudes increase. The cooperative enhancement is even more pronounced when expressed as an intrinsic sample response, which is independent of the sample-to-sensor distance once $\kappa$ is accounted for.  As the feedback approaches the $J$-oscillation threshold, the peak susceptibility increases from approximately 10 without feedback to approximately 80 with feedback gain of $g=-2.2$. Thus, the characteristic molecule-generated field
$\mu_0 M(t)$ reaches approximately $80 \cdot B_a$ given $B_a$ the input magnetic field amplitude. Despite the geometric coupling significantly limits
the corresponding field delivered to the OPM to
$\kappa|\chi|B_a\simeq3.5B_a$ for $\kappa=0.044$, this field now still dominates over the directly applied input field.

\par The frequency-dependent amplification is shown in Fig.\,4e for several feedback gains. As the magnitude of the feedback gain is increased, the resonance becomes progressively stronger and narrower, with the peak amplification reaching approx. 4 for $g=-2.2$. Thus, the feedback enhances the intrinsic response of the molecular medium well beyond
its limit when no feedback is applied. Simultaneously, the resonance lineshape evolves from the dispersive profile characteristic of the intrinsic molecular response ($g=0$) toward a predominantly absorptive peak. This evolution reflects the increasing dominance of the molecule-generated magnetic field over the directly applied input field. For the experimentally chosen delay $t_d=222\,\mathrm{ms}$, the observed amplification is described by
\begin{equation}\label{eq:Ag}
A(\omega,g)=\left|1 +
\frac{g_{3/2} \Gamma_{3/2}}
{(4\pi J-\omega)+i \Gamma_{3/2}(1+g_{3/2}g)}
\right| \, ,
\end{equation}
which predicts both the linewidth narrowing and the enhanced peak amplification as the feedback gain approaches the threshold for $J$-oscillation. The solid curves in Fig.\,4e are calculated from Eq.\,(\ref{eq:Ag}), with $\Gamma_{3/2}$ and $g_{3/2}$ fixed to the values extracted from the linear fit in Fig.\,3c. The resulting theory curves are in excellent agreement with the measured amplification spectra.

\par To further test the molecular media's gain response, Figure\,4f plots the amplification as a function of feedback gains for two excitation frequencies located slightly below and above the $2J$ resonance (marked as dashed vertical lines in Fig.\,4b). The measured amplification is in excellent agreement with the prediction of Eq.\,\ref{eq:Ag} evaluated at the corresponding fixed frequencies, using the same values extracted from Fig.\,3c. As the feedback gain approaches the $J$-oscillation threshold, the amplification increases rapidly owing to the feedback-induced reduction of the effective relaxation rate. As the feedback gain approaches the $J$-oscillation threshold, the magnitude of both the real and imaginary components of $\chi$ increases substantially. The imaginary component remains positive at both frequencies, whereas the real component changes sign across the resonance, reflecting the opposite frequency detunings of the two measurements.

\begin{figure}[tbp]
    \centering
\includegraphics[width=\linewidth]{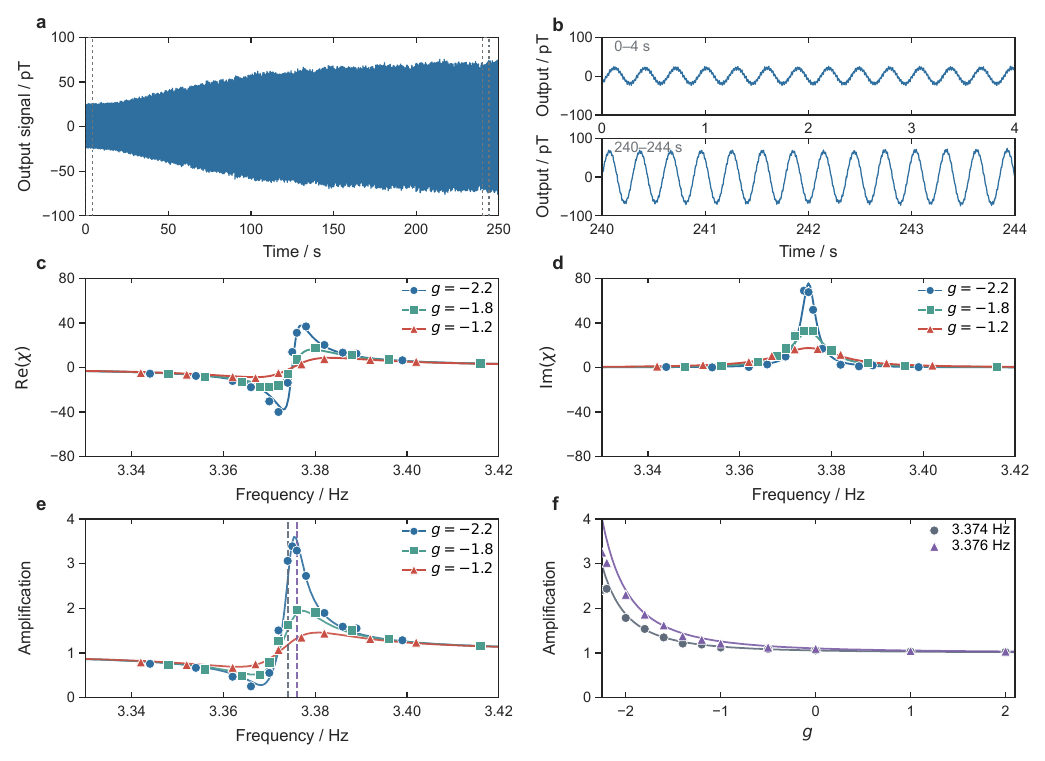}
    \caption{(a) Time-domain traces of the molecular-amplifier output field recorded by the OPM under active feedback with gain $g=-2.2$ and delay $\tau=222$\,ms. (b) Zoomed-in traces from (a) at the beginning (0--4\,s) and near steady state (240--244\,s). (c,d) Real and imaginary components, respectively, of the frequency-dependent susceptibility measured at feedback gains $g=-2.2$, $-1.8$, and $-1.2$. (e) Frequency-dependent amplification for the same feedback gains. Dashed vertical lines mark the excitation frequencies 3.374\,Hz (slate) and 3.376\,Hz (purple). (f) Amplification as a function of feedback gain $g$ at fixed excitation frequencies of 3.374\,Hz (slate) and 3.376\,Hz (purple). Markers denote measurements; solid lines are theoretical predictions using the calibrated internal gain and intrinsic relaxation rates extracted from Fig.\,3(c).}
    \label{fig:cooperative}
\end{figure}

\section{Conclusion and outlook}
In summary, we have realized a continuous cooperative molecular magnetic amplifier based on SABRE-hyperpolarized $J$-coupled nuclear spins. Hyperpolarization significantly enhances molecular magnetic susceptibility and provides intrinsic magnetic gain; an external active feedback extends the spin coherence, and, thereby, enhances the susceptibility even further. The intrinsic response reaches $|\chi|\simeq10$ without feedback and $|\chi|\simeq80$ with feedback near the $J$-oscillation threshold, resulting in fourfold amplification of a $20\,\mathrm{pT}$ input as probed by a nearby optically pumped magnetometer (OPM).

\par The difference between the intrinsic susceptibility and the measured amplification also identifies a clear route toward improved performance. In the present geometry, only the fraction ($\kappa = 0.044$) of the molecule-generated field is delivered to the OPM.  The measured gain is therefore limited primarily by sample-to-sensor distance. Increasing $\kappa$---for example, by placing a compact OPM vapor cell or a solid-state sensor, such as an NV-center diamond magnetometer, directly adjacent to the sample or increasing the polarized sample volume---would convert a larger fraction of the intrinsic molecular response into a detected field.

\par As in other resonant amplifiers, operation closer to threshold compromises bandwidth and response time for gain, with the enhanced susceptibility accompanied by a narrower linewidth.
The operating frequency can be tuned over a wide range through molecular design and isotopic substitution, while finer \textit{in situ} tuning can be achieved by adjusting the feedback delay and/or applying weak bias magnetic fields to shift/split the $J$-transitions. Hyperpolarized molecular ensembles may therefore provide frequency-selective gain media for hertz-frequency weak-field measurements, with potential applications in biomagnetism, geomagnetic sensing, and searches for ultralight dark matter.

Beyond magnetic amplification, continuous measurement of the complex susceptibility provides a complementary approach to zero-field molecular spectroscopy. By sweeping a weak phase-referenced magnetic field across the frequency range of interest and measuring the in-phase and quadrature response, both $\mathrm{Re}[\chi(\omega)]$ and $\mathrm{Im}[\chi(\omega)]$ can be obtained continuously. The resulting spectra provide access to transition frequencies, linewidths, relative spectral amplitudes, and the sign of the nonequilibrium spin order. Unlike conventional free-decay zero-field NMR, continuous SABRE pumping replenishes the spin order throughout the measurement, enabling sustained acquisition and averaging of molecular spectra.

\par More broadly, the cooperative-amplification principle is not restricted to $J$-coupled systems at zero field. Radiation damping at high-field NMR
induces a current in a resonant detection coil which acts as a
feedback to the spin ensemble. The resonator quality factor
$Q$ plays a role analogous to the external feedback gain used in this work. Therefore, equations derived in this work can be also applied to describe the behavior of rasers (radiofrequency amplification via stimulated emission of radiation) \cite{szoke1958radiation,mao1994lineshape,krishnan2013radiation,altafini2010feedback,chacko2024multimode}. For a population-inverted ensemble, the effective transverse relaxation rate takes the analogous form $\Gamma_{\rm eff}=\Gamma_2(1+g_{\rm in}Q)$, with $g_{\rm in}<0$. Increasing $Q$ therefore progressively reduces the effective decoherence and enhances the driven magnetic susceptibility. Since the threshold condition is $|g_{\rm in}|Q=1$, a
higher-$Q$ resonator requires a smaller intrinsic spin gain to reach the same degree of cooperative enhancement. Thus, even a molecular ensemble  with modest internal gain could exhibit substantial amplification when coupled to a sufficiently high-$Q$ resonator. This extension could bring cooperative molecular amplification from the hertz regime to tens or several hundred megahertz. Hyperpolarized molecular ensembles, therefore, provide a broadly tunable class of active, phase-coherent gain media for magnetic sensing and precision measurement.

\clearpage
\section*{Methods}
\subsection*{Sample preparation}
The sample was prepared by mixing $50~\mu\mathrm{L}$ of isotopically
enriched $[^{15}\mathrm{N}]$-acetonitrile with $950~\mu\mathrm{L}$
of acetonitrile-$d_{3}$, yielding a $5\%$ labeled sample.
The mixture contained $12.8~\mathrm{mg}$ of the iridium SABRE
precatalyst and $12.8~\mu\mathrm{L}$ of benzylamine
(BnNH$_2$). During hyperpolarization and signal acquisition,
parahydrogen was continuously bubbled through the sample at a flow
rate of $200~\mathrm{scc/m}$.
\section*{Funding}
This work was supported in part by the Alexander von Humboldt Foundation in the framework of the Sofja Kovalevskaja award, in part by the Deutsche Forschungsgemeinschaft (DFG, German Research Foundation) in the framework of the collaborative research center ``Defects and Defect Engineering in Soft Matter''
(SFB1552) under Project No. 465145163 and under the DFG/ANR grant BU 3035/24-1. D.A.B. acknowledges the University of Miami Provost's research award.

\section*{Author contributions}
Conceptualization: J.X., R.K., M.J., X.P., D.B., D.A.B.\\
Supervision: D.B., D.A.B.\\
Data acquisition: J.X. \\
Theoretical derivations and simulations: J.X. \\
Writing -- original draft: J.X., D.A.B. \\
Writing -- review $\&$ editing: J.X., R.K., M.J., X.P., D.B., D.A.B.

\section*{Competing interests}
The authors declare no competing interests.

\section*{Data availability}
The data that support the findings of this study are available from the corresponding author upon reasonable request.

\section*{Code availability}
The custom code used for data analysis and numerical simulations is available from the corresponding author upon reasonable request.

\bibliographystyle{sn-nature}
\bibliography{ref}

\end{document}